\documentclass[12pt]{article}
\usepackage{jheppub}

\usepackage{ytableau}
\usepackage[export]{adjustbox}
\usepackage{lscape}

\usepackage{amsmath}
\usepackage{axodraw2}

\usepackage{mathabx}

\usepackage{color}

\definecolor{myZero}{rgb}{0.858, 0.188, 0.478}
\definecolor{LinearBox}{RGB}{51, 102, 0}
\definecolor{LtGr}{RGB}{100,252,100}
\usepackage{exscale,latexsym}
\usepackage{amsmath}

\usepackage{cite}

\def\Year{\expandafter\eatPrefix\the\year}
\newcount\hours \newcount\minutes
\def\monthname{\ifcase\month\or
January\or February\or March\or April\or May\or June\or July\or
August\or September\or October\or November\or December\fi}\def\shortmonthname{\ifcase\month\orx
Jan\or Feb\or Mar\or Apr\or May\or Jun\or Jul\or
Aug\or Sep\or Oct\or Nov\or Dec\fi}

\def\TimeStamp{\hours\the\time\divide\hours by60%
\minutes -\the\time\divide\minutes by60\multiply\minutes by60%
\advance\minutes by\the\time%
${\rm \shortmonthname}\cdot   \if\day<10{}0\fi\the\day\cdot   \the\year
\qquad\the\hours:\if\minutes<10{}0\fi\the\minutes$}

\newskip\humongous \humongous=0pt plus 1000pt minus 100pt

\newif\ifdtup

\newcounter{eqnumber}[section]

\def\sset#1{#1}

\def\spa#1.#2{\left\langle#1\,#2\right\rangle}
\def\spb#1.#2{\left[#1\,#2\right]}

\def\tr{\mathop{\rm Tr}\nolimits}

\newbox\charbox
\newbox\slabox
\def\s#1{{      
        \setbox\charbox=\hbox{$#1$}
        \setbox\slabox=\hbox{$/$}
        \dimen\charbox=\ht\slabox
        \advance\dimen\charbox by -\dp\slabox
        \advance\dimen\charbox by -\ht\charbox
        \advance\dimen\charbox by \dp\charbox
        \divide\dimen\charbox by 2
        \raise-\dimen\charbox\hbox to \wd\charbox{\hss/\hss}
        \llap{$#1$}
}}

\def\spa#1.#2{\langle#1\,#2\rangle}
\def\spb#1.#2{[#1\,#2]}
\def\lor#1.#2{\left(#1\,#2\right)}

\def\spba#1.#2.#3{[ #1  | K_{#2} | #3 \ra  }
\def\spaa#1.#2.#3.#4{\la #1 | K_{#2} K_{#3} | #4 \ra }
\def\s#1{s_{#1}}

\catcode`@=11  

\def\Tr{\, {\rm Tr}}

\def\la{\langle}
\def\ra{\rangle}

\def\lsl{\not{\hbox{\kern-2.3pt $\ell$}}}
\def\ksl{\not{\hbox{\kern-2.3pt $k$}}}

\def\tr{\mathop{\hbox{\rm Tr}}\nolimits}

\title{Color Decompositions of the Two Loop Amplitudes of Yang-Mills theory}

\author[]{David~C.~Dunbar}

\affiliation[]{Department of Physics,\\
Faculty of Science and Engineering, \\
Swansea University,\\
Swansea, SA2 8PP, UK }

\emailAdd{d.c.dunbar@swansea.ac.uk}

\abstract{The color structure of two-loop gluon amplitudes is examined both from a color trace basis expansion and an alternative based upon structure constants.
We use use this as a vehicle for systemising relations between the partial amplitudes of the color trace formalism.}


\begin{document}

\maketitle

\def\BB{B}
\def\BA{A}
\def\BC{C}
\def\TB{b}
\def\TA{a}
\def\TC{c}
\def\K{P}

\section{Introduction}

The scattering amplitudes of gluons within a pure $SU(N_c)$ gauge theory are important from a phenomenological point of view
where there is considerable demand for new predictions, particularly in
``Next-to-Next-to-Leading Order'' (NNLO)~\citep{Amoroso:2020lgh}.  
Also,  amplitudes are the custodians of the symmetries 
of the theory  and as such are  important theoretical objects encapsulating information on the symmetries and properties of a theory making their study valuable in itself.

The amplitudes for gluon scattering are functions of both the kinematic variables of the scattered particles but also depend upon their gauge charges or color.   The Feynman diagrams for gluon scattering are generated from the three- and four-point Feynman vertices
\begin{eqnarray*}
V_3 &=& g f^{abc} \left(  \eta_{\mu\nu} ( k_1-k_2)_\lambda  + \eta_{\nu\lambda} ( k_2-k_3)_\mu +\eta_{\lambda\mu} ( k_3-k_1)_\nu  \right)  
\\
V_4 &=& 
  i g^2 \left( f^{acx}f^{bdx} \left( \eta_{\mu\nu}   \eta_{\lambda\delta} -\eta_{\mu\delta}   \eta_{\nu\lambda} \right)  
+f^{adx}f^{bcx} \left( \eta_{\mu\nu}   \eta_{\lambda\delta} -\eta_{\mu\lambda}   \eta_{\nu\delta} \right) 
+ f^{abx}f^{cdx} \left( \eta_{\mu\lambda}   \eta_{\nu\delta} -\eta_{\mu\delta}   \eta_{\nu\lambda} \right) \right)
\end{eqnarray*}   
where  $f^{abc}$ are the structure constants of the group taken to be fully antisymmetric and $k_i$ is the momentum of the $i$-th gluon. 
Any diagram thus combines  color and kinematics.
Although there are three and four point vertices it is a naive observation that the structure constants $f^{abc}$ are purely cubic.
The full amplitude can be expanded in terms of color structures $C^\lambda$ multiplying partial amplitudes that contain the kinematic dependence.
Specifically, we can expand the $n$-point, $\ell$-loop amplitude
\begin{equation}
{\cal A}_n^{(\ell)} = \sum_{\lambda} { A}_{n:\lambda}^{(\ell)} C^\lambda\, . 
\end{equation}
The $C^\lambda$ will be some combination of products of the structure constants.    Ideally, we wish to choose the set of $C^\lambda$ to be a basis.   
This article will be about the structure of two-loop amplitudes under two choices of $C^\lambda$: namely a color trace basis and a basis formed from a subclass of products of structure constants. For tree level and one-loop amplitudes these are well understood and have proven useful in understanding relations between partial amplitudes~\citep{DelDuca:1999rs}.
Most previous analytic results for two-loop gluon scattering have used a color trace basis and organised amplitudes according to the external helicity of the gluons.

\begin{figure}[ht]
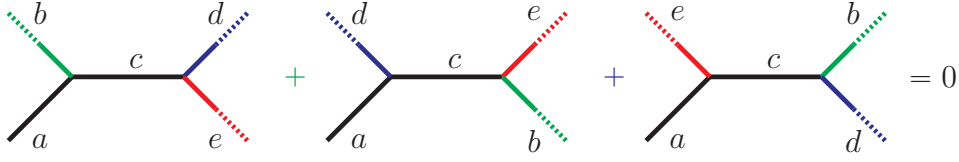

\SetScale{0.6}
    \begin{axopicture}(200,100)(-50,0)   
\Text(-10,10)[c]{$a$}
\Text(-10,90)[c]{$b$}
\Text(100,10)[c]{$e$}
\Text(100,90)[c]{$d$}
\Text(50,60)[c]{$c$}
\SetWidth{3} 
\Line(10,50)(80,50)
\Line(10,50)(-30,10)
\SetColor{Blue}
\Line(80,50)(100,70)
\DashLine(100,70)(120,90){2}
\SetColor{Red}
\Line(80,50)(100,30)
\DashLine(100,30)(120,10){2}
\SetColor{Green}
\Line(10,50)(-10,70)
\DashLine(-10,70)(-30,90){2}
\Text(150,50)[c]{$+$}
 %
 %
  \end{axopicture} 
   \begin{axopicture}(150,100)(-50,0)   
   \Text(-10,10)[c]{$a$}
\Text(-10,90)[c]{$d$}
\Text(100,10)[c]{$b$}
\Text(100,90)[c]{$e$}
\Text(50,60)[c]{$c$}
\SetWidth{3} 
\Line(10,50)(80,50)
\Line(10,50)(-30,10)
\SetColor{Red}
\Line(80,50)(100,70)
\DashLine(100,70)(120,90){2}
\SetColor{Green}
\Line(80,50)(100,30)
\DashLine(100,30)(120,10){2}
\SetColor{Blue}
\Line(10,50)(-10,70)
\DashLine(-10,70)(-30,90){2}
\Text(150,50)[c]{$+$}
 %
 %
  \end{axopicture} 
  \begin{axopicture}(150,100)(-100,0)   
\Text(-10,10)[c]{$a$}
\Text(-10,90)[c]{$e$}
\Text(100,10)[c]{$d$}
\Text(100,90)[c]{$b$}
\Text(50,60)[c]{$c$}
\SetWidth{3} 
\Line(10,50)(80,50)
\Line(10,50)(-30,10)
\SetColor{Green}
\Line(80,50)(100,70)
\DashLine(100,70)(120,90){2}
\SetColor{Blue}
\Line(80,50)(100,30)
\DashLine(100,30)(120,10){2}
\SetColor{Red}
\Line(10,50)(-10,70)
\DashLine(-10,70)(-30,90){2}
\SetColor{Black}
\Text(150,50)[c]{$= 0$}
 %
 %
  \end{axopicture} 
    \caption{Diagramatic representation of the Jacobi Identity}
    \label{fig:Jacobi}
\end{figure}

Any product of structure constants,
\begin{equation}
f^{abc} f^{def}\cdots   f^{xyz}  \; , 
\end{equation}
can be expressed as a ``color diagram'' with cubic vertices
where each structure constant corresponds to an ordered vertex and a repeated index (which conventionally implies summation) indicates a line joining these vertices. The free indices correspond to external gluons.  The set of color diagrams is exactly the set of Feynman diagrams using only the three point vertex. 
This color factor will receive contributions from the corresponding Feynman diagram but also from diagrams with four-point vertices. 
However, these color factors are not all independent because of the Jacobi identity,
\begin{equation}
f^{abc}f^{cde}  +f^{adc}f^{ceb}  +f^{aec}f^{cbd}  = 0 \; , 
\end{equation}  
shown schematically in fig.~\ref{fig:Jacobi}.
This implies that the set of all combinations of structure constants is not a good basis for a color decomposition. 

Following,  Del Duca, Dixon  and Maltoni \citep{DelDuca:1999rs}, 
the Jacobi identity can be used to re-express any combination of structure constants arising from a tree Feynman diagram as a  sum of ``chains'', $Ch_{1,n}$ where \citep{DelDuca:1999rs}
\begin{equation}
Ch_{1,n}(a_2,a_3\cdots , a_{n-1}) \equiv f^{1a_2b_1}f^{b_1a_3b_2} \cdots f^{b_{n-3}a_{n-1}n} 
\end{equation}
where two legs $1$ and $n$ are selected. The $a_i$ are the other external legs and the $b_i$ are repeated indices. 
Diagrammatically, this as shown in fig.~\ref{fig:Chain}.
\begin{figure}[ht]
    \begin{picture}(200,90)(-50,0)   
\SetWidth{2}
\Text(0,20)[c]{1} 
\Text(360,20)[c]{n} 
\Text(60,80)[c]{$a_2$} 
\Text(110,80)[c]{$a_3$} 
\Text(310,80)[c]{$a_{n-1}$} 
\Text(85,30)[c]{$b_1$} 
\Text(135,30)[c]{$b_2$} 
\Text(285,30)[c]{$b_{n-3}$} 
\Line(10,20)(350,20)
\Line(60,20)(60,70)
\Line(110,20)(110,70)
\Line(160,20)(160,70)
\Line(210,20)(210,70)
\Line(260,20)(260,70)
\Line(310,20)(310,70)
\SetColor{Blue}
  \end{picture} 
    \caption{Diagrammatic representation of a ``chain''}
    \label{fig:Chain}
\end{figure}
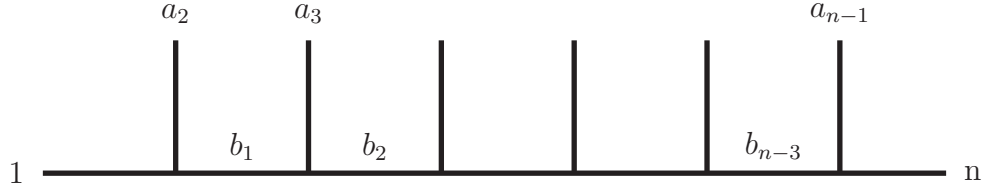 

\begin{figure}[ht]
 \begin{picture}(150,120)(20,0)   
\SetWidth{2} 
\Vertex(40,20){2}
\Vertex(60,20){2}
\Vertex(120,20){2}
\Vertex(100,20){2}
\Line(10,10)(150,10)
\Line(30,10)(30,40)
\Line(130,10)(130,40)
\Line(80,10)(80,40)
\SetColor{Blue}
\Line(80,40)(50,70)
\SetColor{Green}
\Line(80,40)(110,70)
\SetColor{Black}
\Text(160,25){$=$}
\Text(3,10){$1$}
\Text(155,10){$n$}
\Text(50,80){a}
\Text(110,80){b}
  \end{picture} 
    \begin{picture}(150,120)(10,0)  
\Vertex(40,20){2}
\Vertex(55,20){2}
\Vertex(120,20){2}
\Vertex(105,20){2}    
\SetWidth{2} 
\Text(3,10){$1$}
\Text(155,10){$n$}
\Line(10,10)(150,10)
\Line(30,10)(30,40)
\Line(130,10)(130,40)
\SetColor{Blue}
\Line(66,10)(66,50)
\SetColor{Green}\Line(97,10)(97,50)
\SetColor{Black}
\Text(160,25){$-$}
\Text(66,60){$a$}
\Text(97,60){$b$}
 \end{picture} 
    \begin{picture}(125,120)(0,0)   
\SetWidth{2} 
\Vertex(40,20){2}
\Vertex(55,20){2}
\Vertex(120,20){2}
\Vertex(105,20){2}    
\Text(3,10){$1$}
\Text(155,10){$n$}
\Line(10,10)(150,10)
\Line(30,10)(30,40)
\Line(130,10)(130,40)
\SetColor{Green}
\Line(66,10)(66,50)
\SetColor{Blue}
\Line(97,10)(97,50)
\Text(66,60){$b$}
\Text(97,60){$a$}
  \end{picture}     
    \caption{Using Jacobi Identity to transform trees to individual legs. The $a$ and $b$ can denote individual legs or clusters of legs. The end result has clusters of smaller size attached to the line joining $1$ and $n$. The labels $a$ and $b$ can denote single legs or more complicated trees.}
    \label{fig:JacChain}
\end{figure}
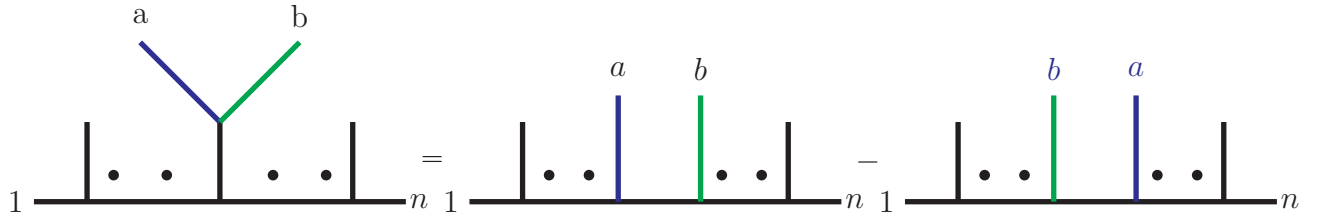 
To see this, once we choose legs $1$ and $n$ then for any color term will have a direct line from $1$ to $n$ with various trees attached.  The Jacobi identity can then be used to split an attached tree into two, smaller, attached trees. This is shown in its simplest form in fig.~\ref{fig:JacChain}. For more complicated configurations, this is repeated until the term is expressed as a sum of chains.
The chains form an independent set and thus a basis for color structures.

Consequently, gathering together the coefficients of the chains, the full amplitude may be expanded~\citep{DelDuca:1999iql,DelDuca:1999rs}, 
\begin{eqnarray}
{\cal A}_n^{(0)}(1,2,3,\cdots ,n)  &=& \sum_{S_{n-2}} 
f^{1a_2b_1}f^{b_1a_3b_2} \cdots f^{b_{n-3}a_{n-1}n}  \overline{A}_{n} (1,a_2,\cdots, a_{n-1} ,  n) , 
\label{eq:decoDDM}
\end{eqnarray}
where since we have selected legs $1$ and $n$ the summation is over the $(n-2)!$ permutations $(a_2,a_3,\cdots, a_{n-1})$ of the remaining legs $2,\cdots , n-1$.

The alternate expansion we consider is
where the color factors are products of the trace of color matrices.  Historically, this existed prior to the expansion in terms of chains. This can be obtained by replacing
\begin{equation}
f^{abc} = \Tr( T^a T^b T^c )-\Tr( T^a T^cT^b)
\end{equation}
where $T^a$ are the generators of the Gauge group.   Using the identities for the traces, 
\begin{equation}
\sum_a \Tr (X T^a ) \Tr (T^a Y)=\Tr(XY)
\;\; , 
\sum_a \Tr (X T^a Y T^a) = \Tr(X) \Tr(Y)
\label{eq:trace_ids}
\end{equation}
the products of traces can be manipulated. At tree level, this reduces to single traces 
and, gathering together contributions,  the full amplitude is
\begin{eqnarray}
{\cal A}_n^{(0)}(1,2,3,\cdots ,n)  &=& \sum_{S_{n}/Z_n}  \Tr[ T^{a_1} T^{a_2} 
\cdots T^{a_n}] A_{n}^{(0)} (a_1,a_2,\cdots, a_n) , 
\label{eq:tree}\end{eqnarray}
in a process which separates color and kinematics.  Note that we can work either with $SU(N_c)$ matrices or $U(N_c)$ matrices. Eqn.~(\ref{eq:trace_ids}) are the $U(N_c)$ identities.  
The partial amplitudes $A_{n}^{(0)} (a_1,a_2,\cdots, a_n)$ are cyclically symmetric but not fully crossing symmetric.
We can use the cyclic symmetry to choose $a_1=1$, then the sum over permutations is over the $(n-1)!$ permutations of $(2,\cdots, n)$.   Additionally,  the partial amplitudes are reflection symmetric 
\begin{eqnarray}
{A}_n^{(0)}(a_1,a_2,a_3,\cdots ,a_n)  &=& (-1)^n  A_{n}^{(0)} (a_n,\cdots, a_3,a_2,a_1)
\; , 
\label{eq:treereflect}\end{eqnarray}
so there are $(n-1)!/2$  individual partial amplitudes.  The above decomposition~(\ref{eq:tree}) can be derived from field theory
~\citep{Cvitanovic:1980bu,Mangano:1990by,Mangano:1988kk}, but naturally arises in string theory
and consequently in field theory~\citep{Kosower:1987ic,Kosower:1988kh} by looking at the infinite string tension limit.

The existence of the two expansions gives useful complementary information.   The color-trace formalism (\ref{eq:tree}) is more symmetric than (\ref{eq:decoDDM}) where two legs have been selected for the expansion.   By considering sufficiently high $N_c$ the external color choices can be chosen so that only a single trace is non-zero.  Consequently, the 
$A_{n}^{(0)}$ are gauge invariant objects.  Additionally, they can be calculated using a color-ordered formalism where only diagrams with the correct ordering are needed~\citep{Mangano:1990by}.

To compare the expressions, we can expand a chain and obtain
\begin{eqnarray}
f^{1a_2b_1}f^{b_1a_3b_2} \cdots f^{b_{n-1}a_{n-1}n} 
&=&\Tr(T^1 T^{a_2}\cdots T^{a_{n-1}} T^{n})+(-1)^n 
\Tr(T^n T^{a_{n-1}} \cdots T^{a_2} T^1)
\notag
\\
&+&\sum \Tr( T^1 X T^n Y) \; , 
\label{eq:f_expansion}
\end{eqnarray}
where the $X$ and $Y$ are non-zero.    Consequently, the only contribution to the coefficient of 
$\Tr(T^1T^{a_2}\cdots T^{a_{n-1}} T^n)$ arising from eqn.~(\ref{eq:decoDDM}) is 
$\overline{A}_{n}(1,a_2 \cdots a_{n-1}, n)$ so we can deduce that
\begin{equation}
\overline{A}_{n}( 1,a_2 \cdots a_{n-1} , n)= 
A_n^{(0)}( 1,a_2 \cdots a_{n-1}, n)  \; , 
\end{equation}
which immediately implies the $\overline{A}_{n}$
are cyclically symmetric. 
In addition, the $A_n^{(0}$ where $1$ and $n$ are not adjacent must be expressible in terms of the 
$\overline{A}_n( 1,a_2 \cdots a_{n-1} , n)$ and consequently in terms of the ${A}_n^{(0)}( 1,a_2 \cdots a_{n-1} , n)$.
This relation is the well known Kleiss-Kuijf relation~\citep{Kleiss:1988ne}
\begin{equation}
A_{n}^{(0)} (1,\{\alpha\}, n, \{\beta\}) =  
(-1)^{|\beta|} \sum_{\sigma \in OP(\alpha ,\beta^T)}  
A_{n}^{(0)}( 1 , \{\sigma\} , n)  \; , 
\label{eq:KK}
\end{equation}
where $\alpha$ and $\beta$ are some sets of the remaining indices i.e. 
$\{\alpha\} =\{ a_2 , \cdots a_p\}$ and $\{\beta\} =\{a_{p+1},\cdots a_{n-1}\}$.   
The summation is over the order permutations of $\alpha$ and $\beta^T$. That is permutations of the union of the sets where the ordering of 
$\alpha$ and $\beta^T$ are preserved.  This relation was known well before the expansion of~\citep{DelDuca:1999rs} but the expansion provides a simple proof of these relations.  The Kleiss-Kuif relations reduce the number of independent amplitudes from $(n-1)!/2$ to $(n-2)!$. This exhausts the relations between tree amplitudes which are linear with constant coefficients and which apply to all helicities.

Relations beyond these exist where the coefficients are algebraic functions of momentum invariants. These can be shown to arise from the color-kinematic duality of ~\citep{Bern:2008qj,Bern:2019prr} which gives rise to extra relations where the coefficients are functions of kinematic variables but independent of helicity configuration~\citep{Bjerrum-Bohr:2009ulz,Kosower:2025inx}.  Our focus will be upon linear relations which are a consequence of group theory. 

\subsection{Decoupling Identities}

Although, the traces in the color trace expansion of the two-loop amplitude are independent from the viewpoint of group theory the partial amplitudes multiplying them
are not. The origin of this is that the amplitudes can be, in principle, calculated via
Feynman diagrams.  Consequently, there are relations between the partial amplitudes.

Some of these relations can be derived from {\it decoupling identities}. These come from the observation that the color trace expansion also applies to $U(N_c)$ gauge theories and if one of the external gluons is taken to be the $U(1)$ 
gluon then the total amplitude must vanish. 
Specifically, if we set
\begin{equation}
T^{a_1} = \frac{1}{\sqrt{N_c}} I_{N_c}    \; ,
\end{equation}
and consider the coefficient of $\Tr[ T^{a_2} T^{a_3} 
\cdots T^{a_n}]$
then we must have
\begin{equation}
0 = \sum_{cyc(a_2,a_3,\cdots a_n)}
A^{(0)}_n(a_1, a_2,a_3,\cdots,  a_n) \; ,
\end{equation}
where the summation is over the $n-1$ cyclic permutations of the legs $a_2, \cdots a_n$.
For small $n \leq 6$ these decoupling identities generate  all relations but for $n > 6$ there are relations which go beyond these and the decoupling identities only imply the Kleiss-Kuijf for $n \leq 6$. 
(see table~\ref{tab:Deco} later.)
Decoupling identities exist at one and two loop also but, again, do not generate all relations.

\subsection{One-Loop Expansion}

At one-loop, the decomposition in terms of color traces contains both single trace terms 
and double trace terms~\citep{Bern:1990ux}
\begin{eqnarray}
{\cal A}_n^{(1)}(1,2,&3& ,\cdots ,n)   = \sum_{S_n/Z_n}  N_c  \Tr[ T^{a_1} T^{a_2} 
\cdots T^{a_n}] A_{n:1}^{(1)} (a_1,a_2,\cdots, a_n)
\notag
\\
+ \sum_{r}   \sum_{P_{n:r}}   &&\hskip -15pt
\Tr[ T^{a_1} 
\cdots T^{a_{r-1}}]  \Tr[ T^{a_{r}} 
\cdots T^{a_n}] A_{n:r}^{(1)} (a_1,a_2,\cdots, a_{r-1} ; a_{r}, \cdots ,a_n) \; .
\end{eqnarray}
The single trace term has a factor of $N_c$ and is thus referred to as the leading-in-color (or planar) contribution. 
The summation over $r$ for a $SU(N_c)$ theory is $r=3,\cdots [n/2]$. For a $U(N_c)$ theory the summation is over 
$r=2,\cdots [n/2]$. The $r=2$  term $A_{n:2}^{(1)} (a_1 ; a_2,\cdots, a_n)$  is the amplitude for color structure 
$\Tr(T^{a_1})\Tr(T^{a_2} \cdots T^{a_n})$.   We refer to partial amplitudes such as $A_{n:2}^{(1)}$ as a $U(N_c)$ specific amplitude.

For the one-loop amplitude the double trace terms are not independent but
can be expressed in terms of the leading~\citep{Bern:1994zx,Bern:1994cg}
\begin{equation}
A_{n:r}^{(1)} (a_1,a_2,\cdots, a_{r-1} ; b_{r}, \cdots , b_n)=(-1)^{r} \sum_{\sigma\in COP\{\alpha\}\{\beta^T\}} 
A_{n:1}^{(1)}  (\sigma )
\label{eq:OneLoopSubLeadingF}
\end{equation}
where $\alpha=\{ a_1 ,\cdots a_{r-1}\}$ and $\beta=\{b_{r} \cdots b_n \}$.   The summation is over the ordered permutations as before, but factoring out equivalent permutations due to cyclic symmetry (see appendix of \citep{Dunbar:2023ayw}  for examples). 
This relation allows the double trace terms to be derived from the leading in color terms only.  This reduces the number of functional forms to be computed in a calculation considerably although the formulae (\ref{eq:OneLoopSubLeadingF}) is not necessarily efficient.
For example, the specific result for the $n$-point all-plus helicity amplitude of ref.\citep{Dunbar:2019fcq}
\begin{equation}
A_{n:r}^{(1)}(1^+,2^+,\cdots, r-1^+ ;  r^+ \cdots n^+) = -2i
\frac{  (K_{1\cdots r-1}^2)^2}{ 
\spa1.2\spa2.3 \cdots \spa{(r-1)}.1   \spa{r}.{(r+1)}  \cdots \spa{n}.{r}  } \; ,    
\end{equation}
is considerably simpler than eqn.~(\ref{eq:OneLoopSubLeadingF}) would suggest. 

\begin{figure}[ht]
   \includegraphics[width= 10.5cm]{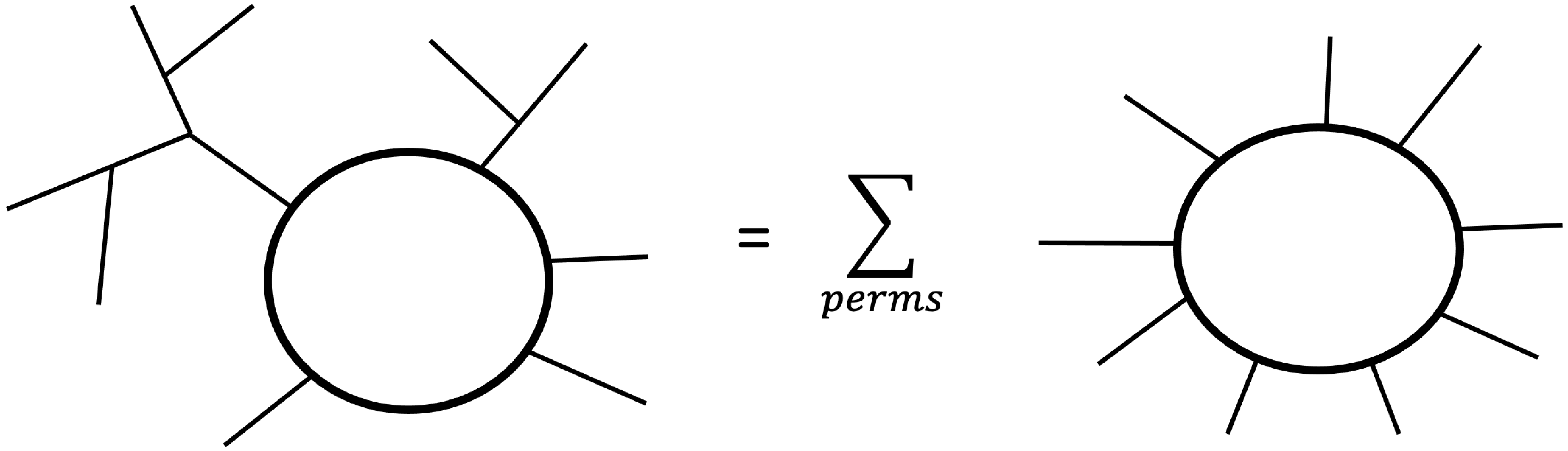}
    \caption{A general one-loop color term can be expressed a sum of cycle terms after repeated application of the Jacobi identity.}
    \label{fig:OneLoopExpansion}
 \end{figure}

If we consider the one-loop equivalent of the expansion in terms of 
structure constants then the general term will correspond to a diagram with a central loop and trees attached to it -such as in the left hand side of fig.~\ref{fig:OneLoopExpansion}.   Using the Jacobi identity repeatedly these can be reduced to  a sum of terms which are a loop with single legs attached as in fig.~\ref{fig:OneLoopExpansion}.    

Thus, at one loop, again following~\citep{DelDuca:1999rs} instead of 
chains we  have ``cycles''
\def\Ri{C^1}
\begin{equation}
\Ri(a_1,a_2,a_3, \cdots , a_n ) \equiv     
f^{b_na_1b_1}f^{b_1a_2b_2} \cdots f^{b_{n-1}a_{n}b_n} \; , 
\end{equation}
which have cyclic symmetry.
Gathering together terms the amplitude can be expanded,
\begin{equation}
{\cal A}_n =  \sum_{S_n/Z_n} \Ri (a_1, a_2, a_3, \cdots , a_n ) 
\overline{A}^{(1)}_n(a_1, a_2, a_3, \cdots , a_n )
\; . 
\end{equation}

Since, the two expansion must be equal we can relate the $\overline{A}^{(1)}_n$ to the $A_{n:r}^{(1)}$ by expanding the $\Ri$ into traces. 
Using eqn.~(\ref{eq:f_expansion}) setting $1=n=b_n$ and relabeling $a_i$, 
\begin{eqnarray}
\Ri(a_1,a_2,a_3, \cdots ,  a_n ) &=& \Tr( T^{a_1} T^{a_2}\cdots T^{a_n} )
+(-1)^n \Tr( T^{a_n} \cdots T^{a_2} T^{a_1} )
\notag\\
&+& \hbox{\rm   double trace terms} \; .
\label{eq:chain_exp}
\end{eqnarray}
We can then equate $\overline{A}_n^{(1)}$ with the leading in color trace term
\begin{equation}
\overline{A}^{(1)}_n(1, a_2, a_3, \cdots , a_n )
={A}^{(1)}_{n:1}(1, a_2, a_3, \cdots , a_n ) \; . 
\end{equation}
Furthermore, clearly the sub-leading in color terms must be related to the $\overline{A}^{(1)}_n$ and hence to the ${A}^{(1)}_{n:1}$.  Detailed enumeration yields the result of 
eqn.~(\ref{eq:OneLoopSubLeadingF}).

\section{Two Loop Basis}

We now turn to the topic of this letter. 
A general two-loop amplitude for the scattering of $n$ gluons in a pure $SU(N_c)$ or $U(N_c)$ gauge theory 
may be expanded in a color trace basis as
\begin{eqnarray}
& & {\cal A}_n^{(2)}(1,2,\cdots ,n) =
N_c^2 \sum_{S_n/\mathcal{P}_{n:1}}  \tr[T^{a_1}T^{a_2}\cdots T^{a_n}] A_{n:1}^{(2)}(a_1,a_2,\cdots ,a_n) \notag \\
&+&
N_c\sum_{r=2}^{[n/2]+1}\sum_{S_n/\mathcal{P}_{n:r} }   \tr[T^{a_1}T^{a_2}\cdots T^{a_{r-1}}]\tr[T^{b_r} \cdots T^{b_n}] 
A_{n:r}^{(2)}(a_1,a_2,\cdots ,a_{r-1} ; b_{r}, \cdots, b_n)  
\notag \\
&+& \sum_{s=1}^{[n/3]} \sum_{t=s}^{[(n-s)/2]}\sum_{S_n/\mathcal{P}_{n:s,t}} 
\tr[T^{a_1}\cdots T^{a_s}]\tr[T^{b_{s+1}} \cdots T^{b_{s+t}}]
\tr[T^{c_{s+t+1}}\cdots T^{c_n}] 
\notag
\\
& & 
\hskip 5.0truecm 
\times A_{n:s,t}^{(2)}(a_1,\cdots ,a_s;b_{s+1} ,\cdots, b_{s+t} ;c_{s+t+1},\cdots, c_n ) 
\notag \\
&+&\sum_{S_n/\mathcal{P}_{n:1}}  \tr[T^{a_1}T^{a_2}\cdots T^{a_n}] A_{n:1B }^{(2)}(a_1,a_2,\cdots ,a_n)\,.
\label{eq:twoloopexpansion}
\end{eqnarray}
The partial amplitudes multiplying any trace of color matrices are cyclically symmetric in the indices within the trace.  The summations simply count each color structure exactly once.    
The expression has single, double and triple trace terms.  In string theory these would arise from surfaces with three boundaries.  
There is also a single trace term $A_{n:1B }^{(2)}$ which is sub-sub leading in powers of $N_c$.  This arises in string theory from a separate two-loop surface with a single boundary~\citep{Dunbar:2020wdh}.

The partial amplitudes are
\begin{equation}
A_{n:r}^{(2)}  \;\;  r=1 \cdots [n/2]   \; , \;\;\  A_{n:r,s}^{(2)}    \;\;  r,s =1 \cdots [n/3],  r \leq s \; , \;\;  A_{n:1B }^{(2)}
\; , 
\end{equation}
of which the $A_{n:2}^{(2)}$ and $A_{n:1, s}^{(2)}$ are the $U(N_c)$ specific functions. 

We now develop an expansion basis for the two-loop amplitude in terms of products of $f^{abc}$.   

\begin{figure}[ht]
    \begin{picture}(300,90)(-50,0)   
\SetWidth{2} 
\Oval(50,45)(30,40)(0)
\Line(50,75)(50,15)
\Oval(150,45)(25,25)(0)
\Oval(220,45)(25,25)(0)
\Line(175,45)(195,45)
\Text(50,0){a)}
\Text(220,0){b)}
\SetColor{Blue}
  \end{picture} 
    \caption{The two possible two loop topologies.  Any color diagram will be consist of one of these with trees of attached.  The Jacobi identity can then reduce these to these topologies with individual legs attached.}
    \label{fig:looptops}
\end{figure}
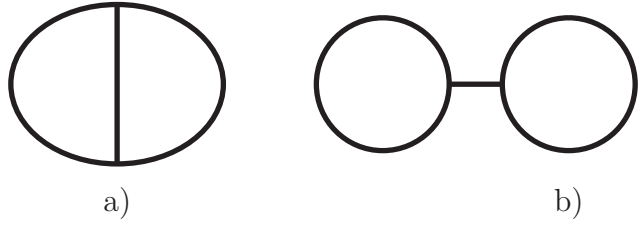

The basic two-loop structure can be of two topologies, where the loop
structures are shown in fig.\ref{fig:looptops}.   Any color term  corresponds to one of these with trees attached to the basis topology.  The first step in creating a basis is to replace, using the Jacobi identity, the trees by individual external legs attached to these two topologies.  Unlike the one loop case,  these is significant further redundancy which we must reduce in order to find a basis. Firstly, as a choice, we can replace all the second, spectacle, topologies, $b)$ ,in terms of the first.

\begin{figure}[ht]
    \begin{picture}(160,90)(-50,0)   
\SetWidth{2} 
\Text(-30,45){$\sset{\alpha}$}
\Vertex(-28,33){1}
\Vertex(-28,57){1}
\Vertex(88,33){1}
\Vertex(88,57){1}
\Oval(0,45)(20,20)(0)
\Line(-14.1,59.1)(-24.1,69.1)
\Line(-14.1,30.9)(-24.1,20.9)
\Oval(60,45)(20,20)(0)
\Line(74.1,59.1)(84.1,69.1)
\Line(74.1,30.9)(84.1,20.9)
\Text(90,45){$\sset{\beta}$}
\Line(20,45)(40,45)
\Line(30,45)(30,65)
\Text(30,70){$a$}
\Text(110,45){$=$}
  \end{picture} 
    \begin{picture}(160,90)(-50,0)   
\SetWidth{2} 
\Vertex(-28,33){1}
\Vertex(-28,57){1}
\Vertex(88,33){1}
\Vertex(88,57){1}
\Text(-30,45){$\sset{\alpha}$}
\Oval(0,45)(20,20)(0)
\Line(-14.1,59.1)(-24.1,69.1)
\Line(-14.1,30.9)(-24.1,20.9)
\Oval(60,45)(20,20)(0)
\Line(74.1,59.1)(84.1,69.1)
\Line(74.1,30.9)(84.1,20.9)
\Text(90,45){$\sset{\beta}$}
\Line(20,45)(40,45)
\Line(43,55)(35,65)
\Text(33,70){$a$}
\Text(110,45){$-$}
  \end{picture} 
      \begin{picture}(110,90)(-50,0)   
\SetWidth{2} 
\Vertex(-28,33){1}
\Vertex(-28,57){1}
\Vertex(88,33){1}
\Vertex(88,57){1}
\Text(-30,45){$\sset{\alpha}$}
\Oval(0,45)(20,20)(0)
\Line(-14.1,59.1)(-24.1,69.1)
\Line(-14.1,30.9)(-24.1,20.9)
\Oval(60,45)(20,20)(0)
\Line(74.1,59.1)(84.1,69.1)
\Line(74.1,30.9)(84.1,20.9)
\Text(90,45){$\sset{\beta}$}
\Line(20,45)(40,45)
\Line(43,35)(35,25)
\Text(33,20){$a$}
  \end{picture} 
    \caption{Identity to remove legs from connecting leg. $\alpha$ and $\beta$ denote an ordered set of legs.}
    \label{fig:IDB}
\end{figure}
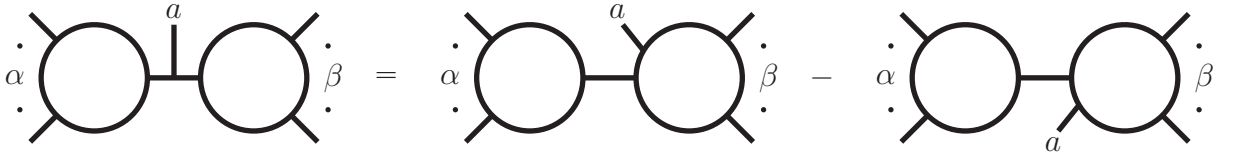

The first step is to eliminate the legs from the line connecting the loops using identity of fig.~\ref{fig:IDB}.  This is used recursively step by step:  starting with $m$ legs this can be expressed as terms with $m-1$ legs which are then expressed as terms with $m-2$ and so on. 
When there are no further legs we can use the identity of fig.~\ref{fig:IDA} to express this as terms of the first type. 
\begin{figure}[ht]
    \begin{picture}(300,90)(-50,0)   
\SetWidth{2} 
\Text(-30,45){$\sset{\alpha}$}
\Oval(0,45)(20,20)(0)
\Line(-14.1,59.1)(-24.1,69.1)
\Line(-14.1,30.9)(-24.1,20.9)
\Oval(60,45)(20,20)(0)
\Line(74.1,59.1)(84.1,69.1)
\Line(74.1,30.9)(84.1,20.9)
\Text(90,45){$\sset{\beta}$}
\Line(20,45)(40,45)
\Text(110,45){$=$}
\Text(130,45){$\sset{\alpha}$}
\Oval(170,45)(20,30)(0)
\Line(170,65)(170,25)
\Text(210,45){$\sset{\beta}$}
\Line(150,59)(140,69)
\Line(150,31)(140,21)
\Line(190,31)(200,21)
\Line(190,59)(200,69)
\Text(225,45){$-$}
\Text(240,45){$\sset{\alpha}$}
\Oval(280,45)(20,30)(0)
\Line(280,65)(280,25)
\Text(320,45){$\overline{\sset{\beta}}$}
\Line(260,59)(250,69)
\Line(260,31)(250,21)
\Line(300,31)(310,21)
\Line(300,59)(310,69)
  \end{picture} 
    \caption{Identity to eliminate spectacle terms}
    \label{fig:IDA}
\end{figure} 

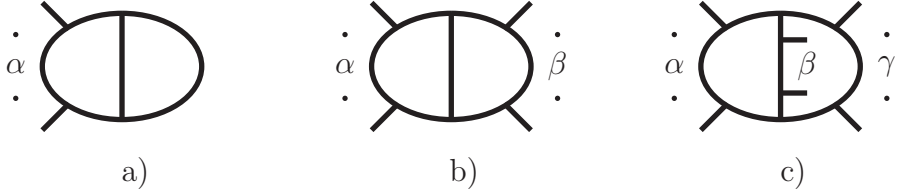
\begin{figure}[ht]
    \begin{picture}(120,90)(0,0)   
\SetWidth{2} 
\Vertex(40,33){1}
\Vertex(40,57){1}
\Text(40,45){$\sset{\alpha}$}
\Oval(80,45)(20,30)(0)
\Line(80,65)(80,25)
\Line(60,59)(50,69)
\Line(60,31)(50,21)
\Text(85,5){a)}
  \end{picture} 
   \begin{picture}(120,90)(0,0)   
\Vertex(40,33){1}
\Vertex(40,57){1}
\Vertex(120,33){1}
\Vertex(120,57){1}   
\SetWidth{2} 
\Text(40,45){$\sset{\alpha}$}
\Oval(80,45)(20,30)(0)
\Line(80,65)(80,25)
\Text(120,45){$\sset{\beta}$}
\Line(60,59)(50,69)
\Line(60,31)(50,21)
\Line(100,31)(110,21)
\Line(100,59)(110,69)
\Text(85,5){b)}  \end{picture}   
  \begin{picture}(120,90)(0,0)   
\Vertex(40,33){1}
\Vertex(40,57){1}
\Vertex(120,33){1}
\Vertex(120,57){1}     
\SetWidth{2} 
\Text(40,45){$\sset{\alpha}$}
\Oval(80,45)(20,30)(0)
\Line(80,65)(80,25)
\Text(120,45){$\sset{\gamma}$}
\Line(60,59)(50,69)
\Line(60,31)(50,21)
\Line(100,31)(110,21)
\Line(100,59)(110,69)
\Line(80,55)(90,55)
\Line(80,35)(90,35)
\Text(90,45){$\sset{\beta}$}
\Text(85,5){c)}
  \end{picture}     
    \caption{Three types of color terms}
    \label{fig:IDC}
\end{figure}

So all color factors can be reduced to a sum of terms of type a) with legs attached to each of the three lines. We could denote this color term by
\begin{equation}
    C^2( \alpha \; ;  \beta \; ;  \gamma )
\end{equation}
where $\alpha$, $\beta$ and $\gamma$ denote an ordered set of external legs, $\alpha=\{a_1,a_2,\cdots a_r\}$ etc.   The $C^2$ can be precisely defining in terms of chains
\begin{equation}
C^2( \alpha \; ;  \beta \; ;  \gamma ) 
=Ch_{x_1y_1}( a_1 \cdots a_r ) 
Ch_{x_2y_2}(b_1 \cdots b_s) 
Ch_{x_3y_3}(c_1 \cdots c_t)  f^{y_1y_3y_2}f^{x_1x_2x_3} 
\end{equation}
which is symmetrical between the three sets. (Note this make diagrams rather clumsy to draw since all legs should be on the same side. We will often not follow this for clarity.)
The color terms can be organised according to the three cases depending upon how many of the $\alpha$, $\beta$ and $\gamma$ are non empty as in fig.~\ref{fig:IDC}

\section{Identities among Color Structures}

The $C^2$ satisfy the following by definition,
\begin{equation}
C^2(\alpha ; \beta ; \gamma  )=   
C^2(\alpha ; \gamma ; \beta  )=   
C^2(\beta ; \alpha ;\gamma  )=   
(-1)^n C^2(\bar\alpha ; \bar\beta ; \bar\gamma  )
\end{equation}
where $\bar\alpha$ is the reversal of set $\alpha$.

There are some general simplifications.  First, consider the color term $a)$ of fig.\ref{fig:IDC}  where all legs lie on a single line, $C^2( \{ a_1, a_2, \cdots a_n  \} ; \{ \} ; \{ \}  )$.  This term naively only has a flip symmetry,
however, evaluating this color term yields $N_c$ times a one-loop color cycle. 
\begin{equation}
       C^2( \{ a_1, a_2, \cdots ,a_n  \};  \{ \} ; \{ \}  ) = N_c \times 
       \Ri(  a_1, a_2, \cdots , a_n  )  \; . 
\end{equation}
From this identity, we can see that this term is actually fully cyclically symmetric and there are only $(n-1)!/2$ independent terms of type a). 
When one of the $\beta$, $\gamma$ is null and the other a single leg, then we have
\begin{equation}
       C^2( \{ a_1, a_2, \cdots a_{n-1}  \}; \{ b_1 \}; \{ \}  ) = -2N_c \times \Ri (   a_1, a_2, \cdots 
       a_{n-1} , b_1 )
\end{equation}
and so these terms are not independent of the a) terms.

\begin{figure}[ht]
    \begin{picture}(300,90)(-50,0)   
\SetWidth{2} 
\SetColor{Red}
\Line(10,72)(0,83)
\Text(10,83){$x$}
\Line(180,60)(193,60)
\Text(200,60){$x$}
\Line(341,73)(351,84)
\Text(340,83){$x$}
\SetColor{Black}
\Vertex(-20,33){1}
\Vertex(-20,57){1}
\Vertex(81,33){1}
\Vertex(81,57){1}
\Oval(30,45)(30,40)(0)
\Line(30,15)(30,75)
\Line(-4,59)(-22,69)
\Line(-4,31)(-22,21)
\Text(-20,45){$\sset{\alpha}$}
\Line(30,40)(44,40)
\Text(50,40){$\sset{\beta}$}
\Line(64,31)(84,21)
\Line(64,59)(84,69)
\Text(85,45){$\sset{\gamma}$}
\Text(100,45){$+$}
\Oval(180,45)(30,40)(0)
\Line(180,15)(180,75)
\Line(146,59)(128,69)
\Line(146,31)(128,21)
\Text(130,45){$\sset{\alpha}$}
\Line(180,40)(194,40)
\Text(200,40){$\sset{\beta}$}
\Line(214,31)(234,21)
\Line(214,59)(234,69)
\Text(235,45){$\sset{\gamma}$}
\Text(250,45){$+$}
\Vertex(130,33){1}
\Vertex(130,57){1}
\Vertex(235,33){1}
\Vertex(235,57){1}
\Oval(330,45)(30,40)(0)
\Oval(330,45)(30,40)(0)
\Line(330,15)(330,75)
\Line(296,59)(278,69)
\Line(296,31)(278,21)
\Text(280,45){$\sset{\alpha}$}
\Line(330,40)(344,40)
\Text(350,40){$\sset{\beta}$}
\Line(364,31)(384,21)
\Line(364,59)(384,69)
\Text(385,45){$\sset{\gamma}$}
\Vertex(280,33){1}
\Vertex(2800,57){1}
\Vertex(385,33){1}
\Vertex(385,57){1}
\Text(410,45){$=0$}
\end{picture}
    \caption{The generic Jacobi identity of eqn.~(\ref{eq_JacobiGen})}
    \label{fig:IDgeneric}
\end{figure}
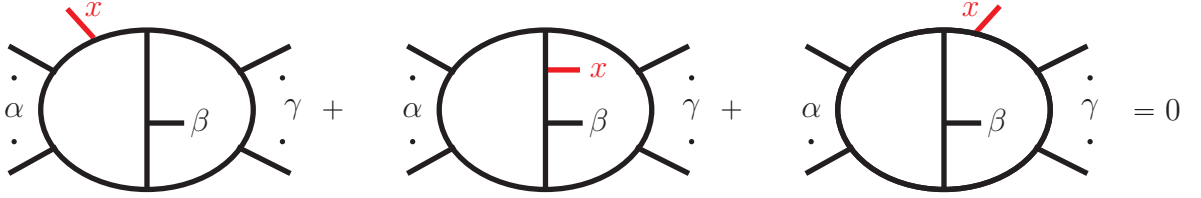

Additionally, the Jacobi identity implies
\begin{equation}
C^2(\alpha \cup \{x\} ; \beta ; \gamma  )+  
C^2(\alpha ; \beta\cup \{x\} ; \gamma  )+ 
C^2(\alpha  ; \beta ; \gamma \cup \{x\} ) =0  
\label{eq_JacobiGen}
\end{equation}
as illustrated in fig.~\ref{fig:IDgeneric}. 
This identity exhausts relations among the $C^2$. 
Repeated application of this identity for example leads to
\begin{equation*}
C^2(\alpha ; \beta ; \gamma  )
=(-1)^{|\gamma|}\sum_{\gamma_1,\gamma_2: OP(\gamma_1,\gamma_2)=\gamma}    C^2( \alpha \cup \overline{\gamma}_1 ; \beta \cup \overline{\gamma}_2  ; \{ \} )
\end{equation*}
where we sum over all partitions of $\gamma$ such that the ordered product regenerates $\gamma$.

The identities of fig.~\ref{fig:IDgeneric}  can be used to radically reduce the spanning set to a much smaller spanning set.  Using the identities, two possible choices of spanning sets are  
\begin{eqnarray}
S_1 &:& \left\{ N_c C^1(\{ a_1,\cdots ,a_n\} ) \;  , 
C^2( \{a_1, \cdots a_r\} ; \{ b_1 \cdots  b_{n-r} \}; \{ \})
\right\}
\notag\\
S_2&:& \left\{ N_c C^1(\{ a_1,\cdots ,a_n\} ) \;  ; 
C^2( \{a_1, \cdots a_r\} ; \{ b_1 \cdots  b_{n-r-1} \}; \{c_1\})
\right\}
\end{eqnarray}
from which a subset will generate a basis.  Note these spanning sets are still significantly larger than a basis.  

We also have the asymmetric spanning sets, 
\begin{eqnarray}
S_3 &:& \left\{ 
C^2( \{ 1, a_1 \cdots a_r\} ;  \{b_1 \cdots  b_{s} \};  \{c_1 \cdots  c_{n-r-s-2}, n\} )
\right\}
\notag
\\
S_4 &:& \left\{ 
C^2( \{ a_1 \cdots a_r\} ;  \{b_1 \cdots  b_{s} \};  \{ 1 \} )
\right\}
\end{eqnarray}
Note that is we consider a $C^2(\alpha;\beta;\gamma)$ where all of $\alpha$, $\beta$ and $\gamma$ are non-zero (described as non-planar) then the expansion into 
color trace terms is
\begin{equation}
C^2(\alpha;\beta;\gamma)= N^2_c.0 +N_c^1 A + N_c^0 B  
\end{equation}
where $A$ and $B$ are some color traces.  That is, these non-planar terms do not contribute to the leading in color $N_c^2$ terms. We also have
\begin{equation}
N_c C^1(a_1,a_2, \cdots a_n)= N^2_c A  +N_c^1 B + N_c^0 .0  
\end{equation}
so this does not contribute to the sub-sub leading terms.

Use of the Jacobi identity indicates relations between the elements of the spanning sets for example as shown in fig.~\ref{fig:IDS1}.

\begin{figure}[ht]
    \begin{picture}(300,90)(-50,0)   
\SetWidth{2} 
%
\Text(-80,45){$\sset{\alpha}$}
\Oval(-40,45)(20,30)(0)
\Line(-40,65)(-40,25)
\Line(-60,59)(-70,69)
\Line(-60,31)(-70,21)
\Line(-20,59)(-10,69)
\Line(-20,31)(-10,21)
\Text(0,45){$\sset{\beta}$}
\Text(-75,21){$x$}
\Text(30,45){$-\sset{\alpha}$}
\Oval(80,45)(20,30)(0)
\Line(80,65)(80,25)
\Line(60,59)(50,69)
\Line(60,31)(50,21)
\Line(100,59)(110,69)
\Line(100,31)(110,21)
\Text(115,21){$x$}
\Text(120,45){$\sset{\beta}$}
\Text(145,45){$+$}
\Text(160,45){$\sset{\alpha}$}
\Oval(200,45)(20,30)(0)
\Line(200,65)(200,25)
\Line(180,59)(170,69)
\Line(180,31)(170,21)
\Line(220,59)(230,69)
\Line(220,31)(230,21)
%
\Text(235,69){$x$}
\Text(240,45){$\sset{\beta}$}
\Text(265,45){$-$}
\Text(280,45){$\sset{\alpha}$}
\Oval(320,45)(20,30)(0)
\Line(320,65)(320,25)
\Text(360,45){$\sset{\beta}$}
\Text(285,69){$x$}
\Line(300,59)(290,69)
\Line(300,31)(290,21)
\Line(340,59)(350,69)
\Line(340,31)(350,21)
\Text(400,45){$=0$}
  \end{picture} 
    \caption{Identity among the $S_1$}
    \label{fig:IDS1}
\end{figure}

In the following sections we will look to determine a basis for the color structures for specific numbers of external legs. It eludes us to find a basis valid in general. Recent work~\citep{Bourjaily:2025hvq} may shed light on this from a formal standpoint. 

\section{Five and Six Point Amplitudes}

At this point it is useful to summarize the results for two-loop gluon scattering  amplitudes.  Results have generally been in the color trace basis and organised according to the helicity of the external gluons. 
The four gluon amplitude was the first  scattering amplitude to   be computed with full-color and for all helicity configurations. This was performed in various dimensional regularisation schemes~\citep{Glover:2001af,Bern:2002tk,Ahmed:2019qtg}. 
Beyond four points progress has been incremental and 
all  partial amplitudes for all helicities of the five gluon amplitude have now been calculated in quite a long series of papers: 
initially the all-plus leading in color partial amplitude~\citep{Badger:2013gxa,Badger:2015lda,Gehrmann:2015bfy,Dunbar:2016aux},
next the remaining all-plus color structures~\citep{Badger:2019djh,Dunbar:2019fcq}
The single-minus helicity leading in color partial amplitude ~\citep{Badger:2018enw} and the remaining leading in color helicity configurations were obtained in~\citep{Abreu:2019odu}, both using finite field numerical methods. Finally, the remaining color structures have been computed~\citep{DeLaurentis:2023nss,Agarwal:2023suw}.

Beyond five point, 
 only the all-plus amplitude has been calculated:
for six gluons, leading in color~\citep{Dunbar:2016gjb}, then full-color~\citep{Dalgleish:2020mof}), using four dimensional unitarity and augmented recursion~\citep{Alston:2012xd}.
and for seven gluon leading in color ~\citep{Dunbar:2017nfy} and full color \citep{Dalgleish:2024sey}.
An $n$-point expression for the all-plus single color trace, $N_c^0$ partial amplitude was conjectured in~\citep{Dunbar:2020wdh}, satisfying various consistency conditions.

These are the specific results against which we may check any relations. 

\subsection{Five point Case} 

If we consider the possible color structures at five points there just four possibilities as shown in table~\ref{tab:my_label}.   
\begin{table}[ht]
    \centering
    \begin{tabular}{c|c|c|c }
    &  term   &  number ${}^1$ & number ${}^2$  \\ \hline
   A &  $C^2 ( \{ a_1, \cdots , a_5 \} ; \{ \} ; \{ \}  )$ & 60 & 12
     \\
     &  $C^2 ( \{ a_1, \cdots , a_4 \} ; \{ b_1 \} ; \{ \}  )$ & 60 & 12
     \\  
    B & $C^2 ( \{a_1,a_2,a_3 \} ; \{ b_1, b_2\} ; \{ \} )$ & 60 & 22 
     \\
  C &   $C^2 (\{ a_1,a_2,a_3\} ; \{b_1\} ; \{c_1\} ) $  & 30 & 10 
         \\
 D &    $C^2 (\{ a_1,a_2 \}; \{b_1, b_2\}; \{c_1\} ) $  & 30 &  10
    \end{tabular}
    \caption{The different Five Point Color Structures with the number of different terms after implementing basic symmetry ${}^1$ together with the number ${}^2$
    of these independent, after Jacobi identities amongst each type. }
    \label{tab:my_label}
\end{table}

The 60 structures of type A can be immediately reduced to the 12 independent cyclic symmetric $N_c\Ri$.  After doing so  the Jacobi identities has rank 110 amongst the 132 color structures indicating that a basis for these has size 22.  
Choices for a basis could be $B'$, $A+B'$, $A+C'$ $A+D'$ where the prime indicates a subset. 
Note that terms of type $A$ only contain $N_c^2$ and $N_c^1$ terms when converted to the trace basis whereas terms of types $C$ and $D$ only contain $N_c^1$ and $N_c^0$. 

A choice of a basis is the 12 cyclic A terms plus the ten independent combinations 
\begin{equation} 
\left\{  C_{ab,cde}   \right\}    
\end{equation}
where
\begin{eqnarray*}
C_{ab,c_1c_2c_3}= 
C^2( \{ c_1,c_2,c_3\} , \{a\}, \{ b \} )  +  
C^2( \{ c_2,c_3,c_1\} , \{a\}, \{ b \} )  +  
C^2( \{ c_3,c_1,c_2\} , \{a\}, \{ b \} )  
\end{eqnarray*}
which satisfies
\begin{equation}
 C_{ab,cde} =C_{ba,cde} =C_{ab,dec} =-C_{ab,ced}    
 \; .
\end{equation}
The amplitude may then be expanded
\begin{equation}
{\cal A}^{(2)}_5 = \sum N_c \Ri(a_1,a_2,a_3,a_4,a_5) F_1(a_1,a_2,a_3,a_4,a_5)+
\sum_{a_1 < a_2 }  C_{a_1 a_2, b_1 b_2 b_3} F_2(a_1,a_2;b_1,b_2,b_3)
\end{equation}
where there are 12 independent $F_1$ and 10 independent $F_2$.

At five points, the color trace basis for $SU(N_c)$ has only three types of partial amplitude
\begin{equation}
A^{(2)}_{5:1}  (12) , \;
A^{(2)}_{5:1B} (12) , \; 
A^{(2)}_{5:3}  (10) , \;
\end{equation}
where the number of independent amplitudes is indicated in brackets. 
Since these can be expressed in terms of the 22 $F_i$  there must be 12 relations among these. 
First note that since the $C_{ab,cde}$ have no $N_c^2$ contribution, comparing expansions
\begin{eqnarray}
A^{(2)}_{5:1} (a_1,a_2,a_3,a_4,a_5)
& =& F_1(a_1,a_2,a_3,a_4,a_5)    \; . 
\end{eqnarray}
 That is, $F_1$ is the leading in color term of the trace basis expansion. 
To proceed further,  we can split the $N_c^1$ term, $A^{(2)}_{5:3}$, into a part arising from the $F_1$ and a part from the $F_2$, 
\begin{eqnarray}
A^{(2)}_{5:3} (a_1,a_2 ; b_1,b_2,b_3) 
& =&  A^{(2),1}_{5:3} (a_1,a_2 ; b_1,b_2,b_3) 
+A^{(2),2}_{5:3} (a_1,a_2 ; b_1,b_2,b_3) 
\end{eqnarray}
where $A^{(2),1}_{5:3}$ is a linear combination of $F_1$ and $A^{(2),2}_{5:3}$ is a linear combination of $F_2$.     The part containing $F_1$,  will arise from the expansion of $\Ri$ in exactly the same way as the one-loop expansion of eqn.~(\ref{eq:OneLoopSubLeading}) and so 
\begin{equation}
A_{5:3}^{(2),1} (a_1,a_2 ; b_{1}, \cdots , b_3)= -\sum_{\sigma\in COP\{\alpha\}\{\beta^T\}} 
A_{5:1}^{(2)}  (\sigma )  \; .
\label{eq:OneLoopSubLeading}
\end{equation}
Expanding the color structure terms $C_{ab,cde}$ and gathering together terms gives
\begin{equation}
A^{(2),2}_{5:3} (a_1,a_2 ; b_1,b_2,b_3) 
=
 -6 F_2(a_1, a_2; b_1,b_2,b_3) 
+ \hskip -0.3 truecm
\sum_{a_1 \leftrightarrow a_2, cyc(b_1,b_2,b_3)}  F_2(a_1,b_1;a_2,b_2,b_3) 
\; , 
\label{eq:A53A}
\end{equation}
and 
\begin{equation}
A^{(2)}_{5:1B} (a_1,a_2,a_3,a_4,a_5)    = -2  \hskip -0.3 truecm
\sum_{cyc(a_1,a_2,a_3,a_4,a_5)} \left(  F_2( a_1,a_2; a_3,a_4,a_5 )-
F_2( a_1,a_3; a_2,a_4,a_5 ) \right) 
\; . 
\label{eq:A51B}
\end{equation}

Equation~(\ref{eq:A53A}) is invertible with
\begin{equation}
36 F_2(a_1, a_2; b_1,b_2,b_3)    =
-7 A^{(2),2}_{5:3} (a_1,a_2 ; b_1,b_2,b_3) 
-\sum_{a_1 \leftrightarrow a_2, cyc(b_1,b_2,b_3)}   A^{(2),2}_{5:3} ({a_1,b_1} ;{a_2,b_2,b_3})   \label{eq:F2}
\end{equation}
However,  eqn.~(\ref{eq:A51B}) is not invertible.  
Equations (\ref{eq:A51B}) and (\ref{eq:F2}) imply that $A^{(2)}_{5:1B}$ may be expressed in terms of 
$A^{(2),2}_{5:3}$.
Specifically, we obtain
\begin{equation}
  A^{(2)}_{5:1B }({a_1,a_2,a_3,a_4,a_5})=\frac{1}{2}\left(
\sum_{cyc(a_1,a_2,a_3,a_4,a_5)} A^{(2),2}_{5:3}({a_1, a_2};{a_3, a_4, a_5}) 
-A^{(2),2}_{5:3}({a_1, a_3}; {a_2, a_4, a_5}) \right)  
\label{eq:five_point_rel}
\end{equation} 
Since $A^{(2),2}_{5:3}=A^{(2)}_{5:3}-A^{(2),1}_{5:3}$ and $A^{(2),1}_{5:3}$ can be expressed in terms of $A^{(2)}_{5:1}$ if follows that $A^{(2)}_{5:1B}$ can be expressed in terms of $A^{(2)}_{5:3}$ and $A^{(2)}_{5:1}$.   

This recovers the result of ref.~\citep{Naculich:2011ep,Edison:2011ta,Edison:2012fn} which was obtained using iteration and color kinematic duality. 
So all relations for the five point color trace amplitudes can be derived using the structure constant expansion.

\subsection{Six Point Case}
For six-point the number of types of color structures grows significantly and is indicated in 
table~\ref{tab:sixpt}.   
\begin{table}[ht]
    \centering
    \begin{tabular}{c|c|c|c }
    &  term   &  number${}^1$ & number${}^2$  \\ \hline
  A &  $C^2 ( \{a_1, \cdots , a_6 \};  \{ \}; \{ \}    )$ & 360 & 60
     \\
  C &   $C^2 (\{a_1,a_2,a_3,a_4\} ; \{b_1, b_2 \}; \{ \} )$ & 360  &  119
     \\
  D &   $C^2 (\{a_1,a_2,a_3\} ; \{b_1, b_2, b_3\} ;\{ \} )$ & 360  &  114
     \\   
  E &    $C^2 (\{a_1,a_2,a_3,a_4\} ; \{b_1 \}; \{c_1\}) $  & 180 &  60
         \\
  F &   $C^2 (\{a_1,a_2, a_3\} ;\{ b_1, b_2\} ; \{c_1\}) $  & 360 &  60
          \\
  G&   $C^2 (\{a_1,a_2\}; \{b_1, b_2\} ; \{c_1, c_2\}) $  & 60 &  60
   \end{tabular}
    \caption{Six Point Color Structures with the number${}^1$ of different terms after implementing basic symmetry together with the number${}^2$ of these independent after Jacobi identities amongst each type. }
    \label{tab:sixpt}
\end{table}
At six-point, the Jacobi  identity has rank 1740 among the 1860 $C^2$ of table~\ref{tab:sixpt} leaving 120
independent.  A  possible basis for the color structure would be to take the 60 independent $A$ terms as $N_c \Ri(a_1,\cdots ,a_6)$ together with the 60 $G$ terms. This combination does indeed forms a basis and so the amplitude can be expanded
\begin{equation}
{\cal A}_6^{(2)} = \sum N_c  \Ri(a, b , c, d , e , f)  F_1( a, b , c, d , e , f )    
+ \sum C^2 ( \{ a, b \} ;\{ c, d\} ; \{e , f \})  F_2 ( a, b ; c, d ; e , f )         
\end{equation}
in terms of two functional forms $F_i$.

The six-point color trace amplitudes are
\begin{equation}
A^{(2)}_{6:1}  (60) , \;
A^{(2)}_{6:1B} (60) , \; 
A^{(2)}_{6:3}  (45) , \;
A^{(2)}_{6:4}  (20) , \;
A^{(2)}_{6:2,2}  (15)  \; .
\end{equation}
Since these must be expressed in terms of the $F_i$,
numerology implies $200-120=80$ relations among these. This is consistent with both the study of the all-plus amplitude and the results of \citep{Edison:2011ta}. It indicates that all relations found from the all-plus also apply to all helicities. 
Decoupling identities allow $A^{(2)}_{6:2,2}$ to be expressed in terms of $A^{(2)}_{6:r}$ so we focus upon the other terms.

Since $F_2$ has no $N_c^2$ terms and from the expansion of $N_c\Ri$, 
\begin{equation}
F_1( a_1, a_2 , a_3, a_4 , a_5 , a_6 )      
= A^{(2)}_{6:1} (a_1, a_2 , a_3, a_4 , a_5 , a_6   )      \; . 
\end{equation}
As for five point, we may split the $N_c^1$   terms in the amplitude into parts linear in the $F_i$
\begin{eqnarray}
A^{(2)}_{6:3} (a_1,a_2 ; b_1,b_2,b_3,b_4) 
& =&  A^{(2),1}_{6:3} (a_1,a_2 ; b_1,b_2,b_3,b_4) 
+A^{(2),2}_{6:3} (a_1,a_2 ; b_1,b_2,b_3,b_4) 
\notag\\
A^{(2)}_{6:4} (a_1,a_2,a_3 ; b_1,b_2,b_3) 
& =&  A^{(2),1}_{6:4} (a_1,a_2,a_3 ; b_1,b_2,b_3) 
+A^{(2),2}_{6:4} (a_1,a_2,a_3 ; b_1,b_2,b_3) 
\notag\\
\label{eq:A6split}
\end{eqnarray}
where $A^{(2),1}_{6:r}$ will be derived from $A^{(2)}_{6:1}$ as in eqn.~(\ref{eq:OneLoopSubLeadingF}). 
Expanding the color structure terms gives
\begin{eqnarray}
A^{(2),2}_{6:3} ( a_1, a_2 ;  b_1, b_2 , b_3 , b_4 ) 
&=&  2  \sum_{cyc(b_1,b_2,b_3,b_4)}
F_2( a_1,a_2 ; b_1,b_2 ; b_4, b_3 )
\\
A^{(2),2}_{6:4} ( a_1, a_2,a_3 ;  b_1, b_2 , b_3  ) 
&=&  - \sum_{cyc(a_1,a_2,a_3)}\sum_{cyc(b_1,b_2,b_3)}\left( 
F_2( a_1,a_2 ; a_3,b_1 ; b_3, b_2 )
+F_2( a_1,a_2 ; b_1,a_3 ; b_3, b_2 ) \right)
\notag
\label{eq:A63}
\end{eqnarray}
and
\begin{eqnarray}
A^{(2)}_{6:1B} ( a_1, a_2,a_3,a_4,a_5 ,& a_6 &  ) 
= 2 F_2^S ({a_1,a_4};{a_2,a_5};{a_3,a_6})
\\&+&
\sum_{cyc(a_1,\cdots a_6)} \left(
       -F_2^S({a_1,a_2} ; {a_3,a_5} ; {a_4,a_6})+
         2 F_2({a_1,a_2} ; {a_3,a_6} ; {a_4,a_5})      
    \right)  
\notag\\&+&
\sum_{a_1 \leftrightarrow a_2}
\left(        -F_2({a_1,a_2};{a_3,a_4} ; {a_5,a_6})
         +2F_2^S({a_1,a_2};{a_3,a_4};{a_5,a_6})
          \right)
 \notag
 \label{eq:A61B}
 \end{eqnarray}
 where
 \begin{equation}
 F_2^S ({a_1,a_2};{a_3,a_4};{a_5,a_6})
 =\sum_{a_3\Leftrightarrow a_4 , a_5\Leftrightarrow a_6}
 F_2 ({a_1,a_2};{a_3,a_4};{a_5,a_6})
 \; . 
 \end{equation}

Equations (\ref{eq:A63}) and (\ref{eq:A61B}) can be inverted to give $F_2$ as a function of $A^{(2)}_{6:1B}$, $A^{(2),2}_{6:3}$ and $A^{(2),2}_{6:4}$ however all  solutions involve all three amplitudes so there is no possibility of a relation such as in eqn.~(\ref{eq:five_point_rel}). 

In summary,  using the color basis based upon structure constants gives the known relations for five and six point amplitudes purely from a group theory analysis.  

An interesting consequence of the decomposition of eqn.~(\ref{eq:A6split}) is  the  the factorisation structure.  As an example, we can examine this for the all-plus amplitude focusing upon the rational terms as computed in ref.~\citep{Dalgleish:2020mof}.  Let $R_{6:x}$ denote the rational terms of $A^{(2)}_{6:x}$ all-plus amplitude.  $R_{6:x}$ contains a wide variety of poles specifically,
\begin{eqnarray}
R_{6:3}( a_1, a_2 ; b_1,b_2,b_3,b_4)  &:& 
\spa{b_i}.{b_{i+1}}^{-2},  
\spa{a_1}.{a_{2}}^{-1}, 
\spa{a_i}.{b_j}^{-1},
t_{b_ib_{i+1}b_{i+2}}^{-1}
\notag 
\\
R_{6:4}( a_1, a_2, a_3 ; b_1,b_2,b_3) &:&
\spa{b_i}.{b_{i+1}}^{-2},  
\spa{a_i}.{a_{i+1}}^{-2}, 
\spa{a_i}.{b_j}^{-1},
t_{b_ib_{i+1}b_{i+2}}^{-1},
t_{a_ia_{i+1}a_{i+2}}^{-1}
 \end{eqnarray}
 where $t_{abc}=(k_a+k_b+k_c)^2$.
For these terms the double pole factors $\spa{b_i}.{b_{i+1}}^{-2}$ make using recursive techniques quite difficult. Recursion techniques~\citep{Britto:2005fq} generally work well when amplitudes have simple poles whose factorisation is understood. If an amplitude contains double poles then both the leading and sub-leading poles must be understood. Although this is possible~\citep{Alston:2012xd,Alston:2015gea} is is considerably more complicated. 
If we split the rational terms $R_{6:x}=R_{6:x}^1+R_{6:x}^2$ as in 
eqn~(\ref{eq:A6split})  then we find that $R_{6:x}^2$ only has simple poles and the double poles are purely contained in  $R_{6:x}^1$ which may be derived from the leading in color $R_{6:1}$.  This opens the possibility of computing the sub-leading partial amplitudes by a combination of a part derived from the leading in color together with a component obtained by straightforward recursion.

\section{Null Vectors and Relations between Color trace Amplitudes}

In refs.~\citep{Naculich:2024fiy} Naculich and Edison studied group color constraints of low point amplitudes but to higher loop level. In doing so they looked at the conversion matrix between two basis for the amplitudes and searched for null vectors of this matrix and in doing so determined the relation s between the color trace amplitudes.  Here we will look at this for higher point amplitudes but specified to two loops.

We can have two color basis : the structure constant basis $C_{S}^i$ and the color trace basis $C_{T}^a$ which depend upon loop level and number of points.   These will be related via
\begin{equation}
 C_{S}^i  = \sum_{a}  M_{i\lambda}  C_{T}^\lambda   
 \label{eq:basischange}
\end{equation}
with the amplitude can be expanded in terms of either
\begin{equation}
{\cal A}_n^{(2)} =  \sum_i C_{S}^i \bar{A}_{n:i}   = \sum_a C_{T}^\lambda A_{n:\lambda}^{(2)}  
\end{equation}

Using (\ref{eq:basischange}) 
\begin{equation}
{\cal A}^{(2)}_n =  \sum_{i,a} M_{i\lambda} C_{T}^\lambda \bar{A}_{n:i}   
= \sum_\lambda C_{T}^\lambda A_{n:\lambda}^{(2)}  
\end{equation}
which implies the relation
\begin{equation}
A_{n:\lambda}^{(2)}= \sum_i  \bar{A}_{n:i}  M_{i\lambda}
\end{equation}

A null-vector of the matrix $M_{i\lambda}$ satisfies
\begin{equation}
\sum_a  M_{i\lambda} V^\lambda  = 0  \;\;\;   \forall i
\label{eq:null_vec}
\end{equation}
So
\begin{equation}
\sum_\lambda A_{n:\lambda}^{(2)} V^\lambda=  \sum_i\sum_\lambda  \bar{A}_{n:i}  M_{i\lambda} V^\lambda = 0     
\label{eq:null_vecB}
\end{equation}
gives an identity among the $ _{n:\lambda}^{(2)}$ and the set of null vectors gives the the set of identities among the color trace amplitudes.  The number of null vectors can be extracted by knowing the rank of the matrix $M_{ia}$.  One of the challenges in this approach is to identify a good basis.

An alternate approach is to start with an expected relation and use this create a
candidate null vector $V^\lambda$ and then check this satisfies 
eqn.~(\ref{eq:null_vec}). It is only necessary to verify this for a spanning set. In ref.~\citep{Dalgleish:2024sey} the all-plus gluon amplitude was studied for seven and eight point and relations among the color trace partial amplitudes determined. Although these relations only apply to the all-plus amplitude they provide candidate relations for all the helicity amplitudes.  The relations were enumerated looking at combinations of amplitudes which form representations under the symmetry group $S_n$ which can be described via Young tableaux.   

\section{Seven and Eight Point Amplitudes}

We now consider higher point amplitudes with the intention to explore whether the expansion can determine the relations between the partial amplitudes in the color trace formalism. 

The seven and eight point amplitudes for the all plus helicity configuration have been studied in \citep{Dalgleish:2024sey} and the linear relations between these determined.  For the seven-point all the partial amplitudes are available. For the eight-point only the cut constructable part of the partial amplitudes was computed.  Additionally, an all-$n$ form of $A^{(2)}_{n:1B}$ was computed via collinear limits~\citep{Dunbar:2020wdh} and unitarity.   It is conjectural as to whether these relations extend to other helicities.  This we aim to address here.

In general, there are more relations than those implied by decoupling identities. 
Specifically for the  sub-sub leading in color amplitude  $A_{n:1B}^{(2)}$ we find
those of table~\ref{tab:Deco}.   A feature of the all-plus is that 
$A^{(2)}_{n:1B}$ satisfies a two-loop version of the Kleiss-Kuijf relation~(\ref{eq:KK}).   
\begin{equation}
A_{n:1B}^{(2)} (1,\{\alpha\}, n, \{\beta\}) =  
(-1)^{|\beta|} \sum_{\sigma \in OP(\alpha ,\beta^T)}  
A_{n:1B}^{(2)}( 1 , \{\sigma\} , n)
\label{eq:KK_two}
\end{equation}
This can be derived from decoupling identities for $n\leq 6$. For $n>6$ it requires further relations. The number of these is shown on table~\ref{tab:Deco} as relation${}^2$. We will examine whether these are a consequence of group theory and as such apply to all helicities. 
\begin{table}[ht] 
\begin{tabular}{c|c|c|c|c|c|}
\hline\hline
n & $\#$  $A_{n:1B}^{(2)}(1,\sigma)$  
& $\#$ $A_{n:1B}^{(2)}(1,\sigma',n)$ & 
relations${}^1$ &  
relations${}^2$ & relations${}^3$  \\
\hline
5 & 12  & 6  & 6 &  0 & 6 \\
6 & 60  & 24 & 36    &  0 & 14 \\
7 & 360 & 120 & 239   & 1 & 36 \\
8 & 2520 &  720 &  1696    &  104 & 685 
\\
\end{tabular}
\caption{The all-plus amplitudes $A_{n:1B}^{(2)}(1,\sigma)$ has a number of relations both among themselves and with other partial amplitudes.  Decoupling identities yield relations${}^1$ only among this amplitude but are valid for all helicities.  The all-plus amplitude exhibits additional relations${}^2$ among the $A_{n:1B}^{(2)}(1,\sigma)$ and also relation between these and the other partial amplitudes${}^3$.}
\label{tab:Deco}
\end{table}

\subsection{Seven point} 

In tables~\ref{tab:sevennum} and~\ref{tab:sevennum_alt} spanning sets for the color structures are shown. A linear analysis in both cases gives the number of independent structures to be $781=360+421$.

\begin{table}[ht]
    \centering
    \begin{tabular}{c|c|c|c }
    &  term   &  no & no indep.  \\ \hline
  A &  $N_c C^1 ( \{ a_1, \cdots a_7    \; \{  \} ;  \{     \}  )$ & 360 & 360
     \\
  C &   $C^2 ( \{ a_1,a_2,a_3,a_4,a_5  \} ;  \{ b_1, b_2   \} ; \{    \}  )$ & 2520  &  781
     \\
  D &    $C^2 ( \{ a_1,a_2,a_3,a_4   \} ;  \{ b_1,b_2,b_3   \} ;  \{    \}) $  & 2520 &  781
         \\
  E &   $C^2 ( \{ a_1,a_2, a_3, a_4,a_5  \};  \{ b_1   \} ;  \{ c_1   \}) $  & 1260 &  421
        \\
  F&   $C^2 ( \{ a_1,a_2,a_3,a_4   \};  \{ b_1, b_2    \};  \{ c_1   \}) $  &  2520&  421
           \\
  G&   $C^2 ( \{ a_1,a_2, a_3   \}; \{  b_1, b_2, b_2    \} ; c_1   \}) $  & 1260 &  421
\\
  H&   $C^2 ( \{  a_1,a_2, a_3   \};  \{ b_1, b_2   \} ;  \{ c_1,c_2) $  & 1260 &  421
   \end{tabular}
    \caption{Seven Point Numerology}
    \label{tab:sevennum}
\end{table}

\begin{table}[ht]
    \centering
    \begin{tabular}{c|c|c|c }
    &  term   &  no & no indep.  \\ \hline
  A &  $C^2 (\{ a_1, \cdots a_6 \} ; \{ 1\}; \{  \}  )$ & 360 & 360
     \\
  B &  $C^2 (\{ a_1, \cdots a_5 \}; \{ b_1 \} ; \{1 \} ) $  & 360 &  331
         \\
  C &   $C^2 (\{ a_1,a_2, a_3, a_4\}; \{ b_1,b_2 \} ; \{ 1 \} ) $  & 360 &  340
        \\
  D &   $C^2 (a_1,a_2,a_3 \}; \{ b_1, b_2, b_3 \}  ; \{ 1 \} ) $  &  180 &  166
   \end{tabular}
    \caption{Alternate smaller but less symmetrical spanning set. $B+C+D$ has number independent of 421 as does $B+C$ and $C+D$}
    \label{tab:sevennum_alt}
\end{table}

In the color trace basis the $1287$ partial amplitudes for $SU(N_c)$ are
\begin{equation}
A^{(2)}_{7:1}  (360) , \;
A^{(2)}_{7:1B} (360) , \; 
A^{(2)}_{7:3}  (252) , \;
A^{(2)}_{7:4}  (210) , \;
A^{(2)}_{7:2,2}  (105) , \;
\end{equation}
Of the 1287 only 781 can be independent, so there must be 506 relations between the trace basis amplitudes. However, the all-plus amplitudes satisfy 521~\citep{Dalgleish:2024sey} so there is a mismatch of 15. 

From the known relations for the all-plus we can form candidate null vectors which can then be checked against the condition of eqn.~(\ref{eq:null_vec}).  It is not necessary to reduce the spanning set to a basis to do so. In ref.~\citep{Dalgleish:2024sey} the relations of the seven point all-plus were determined at the level of representations of the symmetry group~\citep{Edison:2012fn,Dalgleish:2024sey}. Representations of $S_n$ can be characterised by a Young tableaux and there can be multiple realisations of a given representation.  Each term in a representation being a particular sum of $A^{(2)}_{n:\lambda}$  Each relation thus supplies a candidate null vector.   By examining the potential null vectors from these relations we find all yield good null vectors {\it except} that from the relation of eqn.~(\ref{eq:symm_rel}): 
\ytableausetup{centertableaux, boxsize=0.5em}
\begin{equation}
A^{(2)}_{7:4} 
\left(\hbox{\begin{ytableau}  
\null  & \null   &
\null  &  \null &  \null \\  \null \\
\null
\end{ytableau}}
\right)_1
=  -\frac{1}{2}
A^{(2)}_{7:4} 
\left(\hbox{\begin{ytableau}  
\null  & \null   &
\null  &  \null &  \null \\  \null \\
\null
\end{ytableau}}
\right)_2
\label{eq:symm_rel}
\end{equation}
where the subscript refers to the particular copy of the $S_n$ representation. 
Consequently,  all the relations satisfied by the seven point all-plus {\it except} that of eqn.~(\ref{eq:symm_rel}) are a consequence of group theory and as such must be satisfied by all helicity amplitudes.  In particular, the $A^{(2)}_{7:1B}$ amplitude must satisfy the Kleiss-Kuijf like relation (\ref{eq:KK_two}) for all helicities. 

\subsection{Eight Point} 

For the eight point all-plus amplitude only the cut-constructible pieces are known for all the partial amplitudes. Consequently, only these can be used to search for relations in ref.~\citep{Dalgleish:2024sey}.  However, there is one partial amplitude for which both the cut constructible and rational pieces are known : $A^{(2)}_{8:1B} $.  This amplitude does satisfy (\ref{eq:KK_two}). This provides a candidate relations from which to construct a null vector.  Inserting these into (\ref{eq:null_vec}) we find however
\begin{equation}
\sum_a  M_{ia} V^a  \neq  0  \;\;\;   \forall i
\label{eq:null_vectorB}
\end{equation}
Consequently, (\ref{eq:KK_two}) is not a consequence of using a group theoretical basis and may be a specific property of the all-plus helicity amplitude which need not extend to other helicity configurations. 

If we examine the relations organised by the representation of the symmetry group we find the relation of (\ref{eq:eightpt_relA}) which apply to the all-plus, does not generate a valid null vector. 
\begin{equation}
A^{(2)}_{8:1B} 
\left(\hbox{\begin{ytableau}  
\null  & \null   &
\null  &  \null &  \null \\  \null \\
\null\\
\null
\end{ytableau}}
\right)_1
+2
A^{(2)}_{8:1B} 
\left(\hbox{\begin{ytableau}  
\null  & \null   &
\null  &  \null &  \null \\  \null \\
\null\\
\null
\end{ytableau}}
\right)_2
=0 
\label{eq:eightpt_relA}
\end{equation}
Also, one of the relations between the
\begin{equation}
A^{(2)}_{8:1B} 
\left(\hbox{\begin{ytableau}  
\null  & \null   &
\null  &  \null \\  \null &  \null &
\null & \null
\end{ytableau}}
\right)_i    
\end{equation}
do not generate a null vector.  The $A^{(2)}_{8:1B}$ satisfy 55 relations beyond the decoupling but not the 104 needed. 
Consequently, the Kleiss-Kuijf relation (\ref{eq:KK_two}) may not extend beyond seven points for all helicities. 

In ref~\citep{Feng:2011fja} two-loop relations were examined by investigating whether they could be shown to be valid on a triple Unitary cut.   The Unitary cuts suggested the following identities are true for the cut-constructible parts of the amplitudes, 
\begin{eqnarray}
 A^{(2)}_{8:1B} (1,\alpha_1,\alpha_2,8,\beta_1,\beta_2,\beta_3,\beta_4)&+& 
 A^{(2)}_{8:1B} (1,\alpha_2,\alpha_1,8,\beta_1,\beta_2,\beta_3,\beta_4)
 \notag\\
 =  
 \sum_{\sigma\in  OP(\alpha ,\beta^T)} & & A^{(2)}_{8:1B} (1,\sigma,8)+
 \sum_{\sigma\in  OP(\alpha^T ,\beta^T)} A^{(2)}_{8:1B} (1,\sigma,8)
 \notag\\
 \sum_{cyc(\alpha_1,\alpha_2,\alpha_3)} A^{(2)}_{8:1B} (1,\alpha_1,\alpha_2,\alpha_3,8,\beta_1,\beta_2,\beta_3)
 & =& \sum_{cyc(\alpha_1,\alpha_2,\alpha_3)} 
 \sum_{\sigma\in  OP(\alpha ,\beta^T)} A^{(2)}_{8:1B} (1,\sigma,8)
 \notag\\
 \sum_{cyc(\alpha_1,\alpha_2,\alpha_3)} A^{(2)}_{8:1B} (1,\alpha_1,\alpha_2,\alpha_3,\alpha_4,8,\beta_1,\beta_2)
 & &
 \\
 =
 \sum_{cyc(\alpha_1,\alpha_2,\alpha_3)}  & & 
 \sum_{\sigma\in  OP(\{ \alpha,\beta^T)}   
 A^{(2)}_{8:1B} (1,\sigma,8)
 \; .
\notag\end{eqnarray}
These identities would be implied by the Kleiss-Kuijf relations but not vice-versa. 
From these identities, we can form candidate null vectors and test these against eqn.~(\ref{eq:null_vec}).  The result is that they do generate null vectors and hence must be true relations for all helicities and the entire amplitude.

\section{Conclusions} 

The color trace expansion of gauge theory amplitudes is a powerful tool not least because of the use of color ordered formalism to simplify calculations. The partial amplitudes are gauge invariant, have cyclic symmetries and have a relatively direct link with string theory amplitudes. Nonetheless there are inherent redundancies or relations between the various partial amplitudes. The root of this lies in the Feynman diagram expansion in terms of connected diagrams.  These  redundancies lead to well-known relations at tree and one-loop level.  
Here we have used expansion in color terms based upon products of structure constants together with the Jacobi identity to explore alternate expansions of the color structure at two-loop level.  
Although lacking some of the attractive features of the color trace expansion it is possible to use these to determine the number of independent color structures and consequently the redundancy and relations among the two-loop color trace amplitudes.   

Our results essentially recover previous results obtained or suggested via other techniques. In particular, relations which are satisfied by the limited number of available specific calculations. The only mismatch is that the all-plus amplitude appears to satisfy a  number of relations beyond those following from group theory.  It remains unknown whether these are specific to this particular helicity configuration.

This work was partially supported by the UKRI Science and Technology Facilities Council (STFC) Consolidated Grant No. ST/T000813/1.    For the purpose of open access, the authors have applied a Creative Commons Attribution (CC BY) licence.


\bibliography{TwoLoop}{}

\bibliographystyle{JHEP}

\end{document}